\documentclass[twocolumn]{aastex62}

\newcommand{\ha}{\hbox{H$\alpha$}}
\newcommand{\hb}{\hbox{H$\beta$}}

\newcommand{\oiii}{\hbox{[O\,{\sc iii}]}}
\newcommand{\nii}{\hbox{[N\,{\sc ii}]}}

\shorttitle{SGMS of Two Modes}
\shortauthors{Q.Liu et al.}

\usepackage{chngcntr}

\begin{document}

\title{Elevation or Suppression? The Resolved Star Formation Main Sequence of Galaxies with\\ Two Different Assembly Modes}

\author{Qing Liu}
\affiliation{CAS Key Laboratory for Research in Galaxies and Cosmology, Department of Astronomy, University of Science and Technology of China, Hefei 230026, China\\}
\affil{School of Astronomy and Space Science, University of Science and Technology of China, Hefei 230026, China}
\email{lq960823@mail.ustc.edu.cn}

\author{Enci Wang}
\affiliation{CAS Key Laboratory for Research in Galaxies and Cosmology, Department of Astronomy, University of Science and Technology of China, Hefei 230026, China\\}
\affiliation{School of Astronomy and Space Science, University of Science and Technology of China, Hefei 230026, China}
\email{ecwang16@ustc.edu.cn}

\author{Zesen Lin}
\affiliation{CAS Key Laboratory for Research in Galaxies and Cosmology, Department of Astronomy, University of Science and Technology of China, Hefei 230026, China\\}
\affiliation{School of Astronomy and Space Science, University of Science and Technology of China, Hefei 230026, China}

\author{Yulong Gao}
\affiliation{CAS Key Laboratory for Research in Galaxies and Cosmology, Department of Astronomy, University of Science and Technology of China, Hefei 230026, China\\}
\affiliation{School of Astronomy and Space Science, University of Science and Technology of China, Hefei 230026, China}

\author{Haiyang Liu}
\affiliation{CAS Key Laboratory for Research in Galaxies and Cosmology, Department of Astronomy, University of Science and Technology of China, Hefei 230026, China\\}
\affiliation{School of Astronomy and Space Science, University of Science and Technology of China, Hefei 230026, China}

\author{Berzaf Berhane Teklu}
\affiliation{CAS Key Laboratory for Research in Galaxies and Cosmology, Department of Astronomy, University of Science and Technology of China, Hefei 230026, China\\}
\affiliation{School of Astronomy and Space Science, University of Science and Technology of China, Hefei 230026, China}

\author{Xu Kong}
\affiliation{CAS Key Laboratory for Research in Galaxies and Cosmology, Department of Astronomy, University of Science and Technology of China, Hefei 230026, China\\}
\affiliation{School of Astronomy and Space Science, University of Science and Technology of China, Hefei 230026, China}
\email{xkong@ustc.edu.cn}



\begin{abstract}

We investigate the spatially resolved star formation main sequence in star-forming galaxies using Integral Field Spectroscopic observations from the Mapping Nearby Galaxies at the Apache Point Observatory (MaNGA) survey. We demonstrate that the correlation between the stellar mass surface density ($\Sigma_*$) and star formation rate surface density ($\Sigma_{\mathrm{SFR}}$) holds down to the sub-galactic scale, leading to the sub-galactic main sequence (SGMS). By dividing galaxies into two populations based on their recent mass assembly modes, we find the resolved main sequence in galaxies with the `outside-in' mode is steeper than that in galaxies with the `inside-out' mode. This is also confirmed on a galaxy-by-galaxy level, where we find the distributions of SGMS slopes for individual galaxies are clearly separated for the two populations. When normalizing and stacking the SGMS of individual galaxies on one panel for the two populations, we find that the inner regions of galaxies with the `inside-out' mode statistically exhibit a suppression in star formation, {with a less significant trend in the outer regions of galaxies with `outside-in' mode}. In contrast, the inner regions of galaxies with `outside-in' mode and the outer regions of galaxies with `inside-out' mode follow a slightly sublinear scaling relation with a slope $\sim$0.9,  which is in good agreement with previous findings, {suggesting} that they are experiencing a universal regulation without influences of additional physical processes. 

\end{abstract}




\section{Introduction} \label{sec:intro}

One of the well-established relationships in galaxy formation and evolution is the correlation between the star formation rate (SFR) and the stellar mass (M$_*$) for star-forming galaxies (SFGs), which is usually referred to as the star-forming main sequence (SFMS). Observations have demonstrated that the SFMS holds from the local universe (e.g. \citealt{2004MNRAS.351.1151B}; \citealt{2007ApJS..173..267S}) to the high-redshift one (e.g. \citealt{2007ApJ...670..156D}; \citealt{2007A&A...468...33E}; \citealt{2007ApJ...660L..43N}; \citealt{2017ApJ...840...47B}). In addition, the SFMS exhibits an evolution with redshift (summarized in \citealt{2014ApJS..214...15S}).

Recently, a tight correlation between surface stellar mass density ($\Sigma_*$) and surface SFR density ($\Sigma_{\mathrm{SFR}}$) have attracted attention ({e.g. \citealt{2013A&A...554A..58S}}; \citealt{2013ApJ...779..135W}; \citealt{2016ApJ...821L..26C}, hereafter C16; \citealt{2016MNRAS.456.4533M}; \citealt{2017MNRAS.469.2806A}; \citealt{2017ApJ...851L..24H}, hereafter H17; \citealt{2017MNRAS.466.1192M}), which is referred to as the resolved star formation main sequence, or sub-galactic main sequence (SGMS). This indicates a more fundamental relation between local ongoing star-forming activities and the underlying stellar populations. C16 have found an SGMS on an $\sim1$ kpc scale with a slope of $\sim0.7$ in the local universe using Integral Field Spectroscopic (IFS) data from the CALIFA survey \citep{2012A&A...538A...8S}, while other works found a slope approaching unity (0.9--1.0) from local (\citealt{2017MNRAS.469.2806A}; \citealt{2017MNRAS.466.1192M}) to high-redshift (\citealt{2013ApJ...779..135W}; \citealt{2016MNRAS.456.4533M}).

Thanks to IFS surveys, galaxies can be divided into two populations according to their assembly modes (\citealt{2013ApJ...764L...1P}; \citealt{2015ApJ...804L..42P}; \citealt{2016MNRAS.463.2799I}; \citealt{2017MNRAS.466.4731G}). In a parallel paper \citep{2017ApJ...844..144W}, we found that galaxies that are in the recent `outside-in' mode have smaller sizes, higher concentrations, and higher global gas-phase metallicities with respect to galaxies in the recent `inside-out' mode, suggesting that they are likely in the transitional phase from normal SFG to quiescent populations with rapid central stellar mass growth. Thus, it is essential to understand the star formation regulation in these two populations, especially their behaviors on SGMS.  

In this work, we present our results based on $\sim$600 SFGs from the Mapping Nearby Galaxies at the Apache Point Observatory (MaNGA) survey \citep{2015ApJ...798....7B}. The paper is organized as follows: Section \ref{sec_data} describes the sample selection and the calculation of physical parameters. Section \ref{sec_results} presents our main results, including (1) the SGMS of our sample and our subsamples; (2) {the distributions and dependencies of} the SGMS in individual galaxies {for our subsamples}; and (3) the stacking of the galaxy-by-galaxy (G-by-G) SGMS of two assembly modes. Section \ref{sec_discussion} discusses our results with potential explanations. Finally, section \ref{sec_conclusion} comes up with our conclusions. The cosmology parameters adopted are H$_0=$ 71 km s$^{-1}$Mpc$^{-1}$, $\Omega_M=$ 0.27, and $\Omega_{\Lambda}=$ 0.73. 

\section{Data} \label{sec_data}

\subsection{Sample and Classification} \label{sec_gal}
In this work, we use the newly released MaNGA data from SDSS DR14 \citep{2017arXiv170709322A}, including more than 2700 galaxies. The M$_*$, SFR, effective radius (R$\rm _e$), and the other parameters we used are drawn from the MaNGA Pipe3D Value Added Catalog \citep{2017arXiv170905438S}. The sample is first reduced to have an inclination $<$ 60{\arcdeg} to avoid high inclination effects, with $\sim$1850 galaxies left. We use the criteria that NUV-r $<$ 4 to select SFGs \citep{2015ApJ...804..125L}, with $\sim$800 galaxies selected. We further remove galaxies with a field of view (FoV) $<$ 2 R$\rm _e$ to ensure that most of the galaxies in the remaining sample have enough sampling in both inner and outer parts, with 647 galaxies left. {It is worth noting that we are not sampling the entire disk for our sample, which would be a more suitable task for the CALIFA survey \citep{2012A&A...538A...8S} or the MUSE survey \citep{2017A&A...608A...1B}.}

Then we divide our galaxy sample into two populations according to their recent mass assembly modes. Briefly, the classification is based on information from spatially resolved information of the 4000 {\AA} break (D4000) as follows: Galaxies in the `outside-in' mass assembly mode are defined to have $\mathrm{D4000_{1.5R_e}}$ (D4000 at 1.5 R$\rm _e$) $>$ $\mathrm{D4000_{cen}}$ (D4000 within 0.25R$\rm _e$). Conversely, those with $\mathrm{D4000_{1.5R_e}}$ $<$ $\mathrm{D4000_{cen}}$ are classified as being in the `inside-out' mass assembly mode. D4000 is defined as the ratio between the average flux in red-end and blue-end continua at 4000 {\AA} wavelength (\citealt{1983ApJ...273..105B}), which has been widely used to trace the stellar populations with mean stellar ages younger than 1-2 Gyr (e.g., \citealt{1999ApJ...527...54B}; \citealt{2003MNRAS.341...33K}; \citealt{2004MNRAS.351.1151B}; {\citealt{2005MNRAS.362...41G}}; \citealt{2017arXiv171007569W}). More details and discussions about the classification based on D4000 can be found in \cite{2017ApJ...844..144W}.

\begin{figure}
 \includegraphics[width=8.5cm]{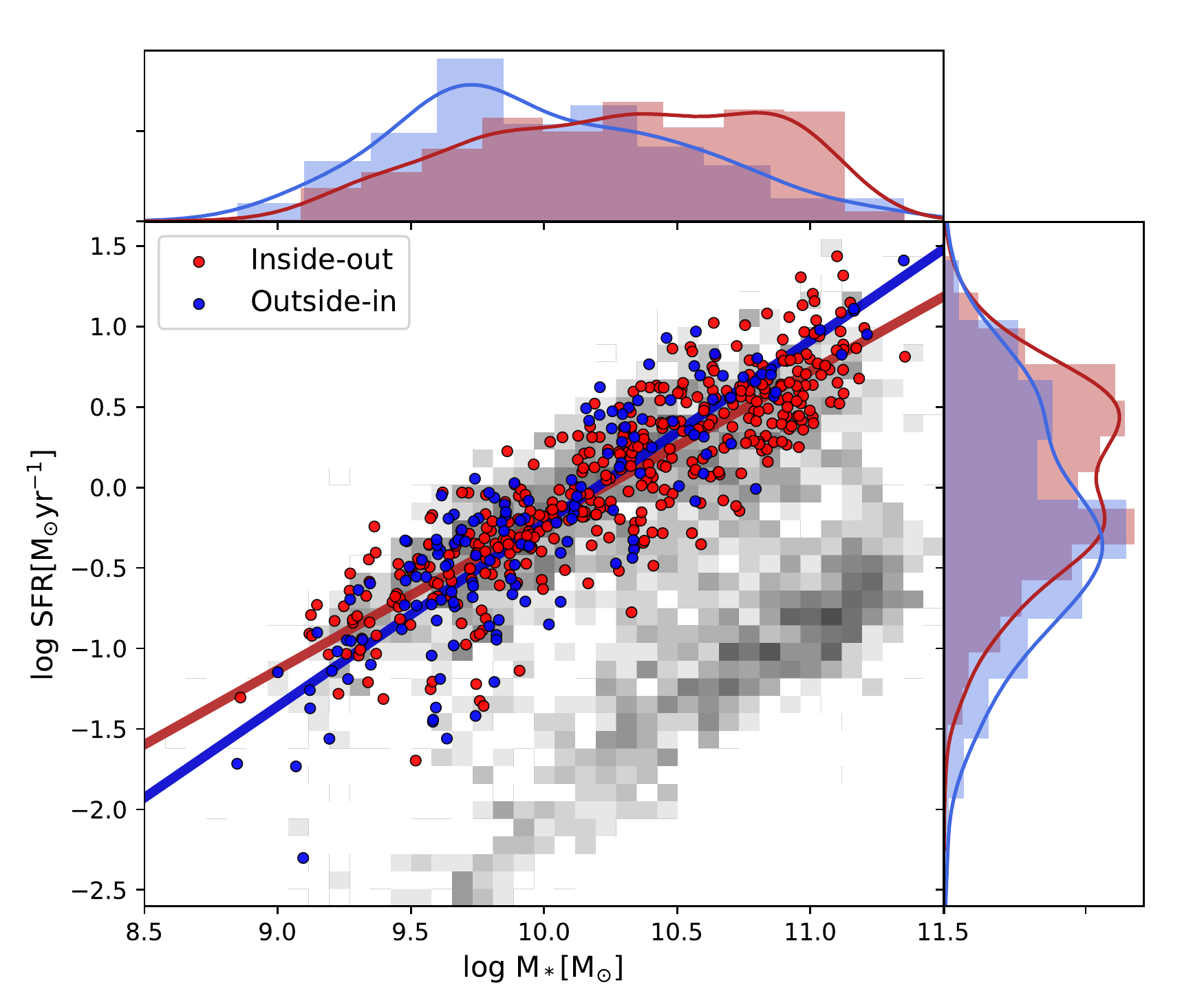}
    \caption{Distribution of sample and subsamples in our study. Red circles stand for `inside-out' galaxies and blue circles for `outside-in' galaxies. Distributions of M$_*$ and SFR for our {subsamples} are shown in small panels by red (`inside-out' galaxies) and blue (`outside-in' galaxies) histograms. Colored lines in red and blue show their linear fittings. {All $\sim$2700 galaxies in the Pipe3D catalog are laid out as the gray background.} On average, massive `outside-in' galaxies are located above `inside-out' galaxies.}
       \label{SFMS}
 \end{figure}

Through visual inspection, we remove major mergers and galaxies with very few available spaxels or dubious R$\rm _e$ measurements for $\sim$7.5\% galaxies in the following analysis. {To avoid resolution effects from the point spread function (PSF) (see Appendix), we remove galaxies with R$\rm _e$ smaller than the PSF (2.5$\arcsec$) of the MaNGA survey, which account for 5\% of the sample}. Finally, we obtain {406} galaxies classified as being in the `inside-out' mass assembly mode and {155} classified as being in the `outside-in' mass assembly mode. For simplicity, we hereafter refer to them as `inside-out' galaxies and `outside-in' galaxies. The distribution of our sample on the $\rm M_*$--SFR plane is shown in Figure \ref{SFMS}. In agreement with \cite{2017ApJ...844..144W}, `outside-in' galaxies $>$ 10$^{10}$ M$_{\odot}$ are typically preferred to be located above `inside-out' galaxies. {We use the same criteria in \cite{2017arXiv170905438S} that their central 2.5 arcsec/aperture regions have (a) emission line ratios above the Kewley limit \citep{2001ApJ...556..121K} and (b) EW(\ha) $>$ 3 {\AA}{\ }, to classify the AGN hosts in our sample and find an AGN ratio to be 1.7\% in `inside-out' galaxies and 2.6\% in `outside-in' galaxies. However, these ratios should not be overemphasized, due to the limited sample size.}

{The 25\%--50\%--75\% quantiles of redshift distributions for our subsamples are 0.027--0.040--0.053 (`inside-out') and 0.022--0.026--0.037 (`outside-in'). Therefore, our following analysis should represent results 0.35--0.7 Gyr ago for `inside-out' galaxies and 0.3--0.5 Gyr ago for `outside-in' galaxies. Considering the dependence of M$_*$  and range of local densities with redshift in the MaNGA sample, we perform tests by restricting galaxies to redshift $<$ 0.04 and find that the main results do not change. Therefore, the slight redshift difference of two subsamples would not influence our conclusions.} 

{On the other hand, due to the target selection strategy, the MaNGA sample is not a good representative sample without a volume correction. Therefore, we perform a volume correction on the numbers of the subsamples, which results in a ratio of 66.5\% for inside-out galaxies and 33.5\% for outside-in galaxies. The color-enhanced subsample cannot be volume-corrected properly; though, it accounts for a very small portion in our sample ($<$2\%) and would not have significant influence on our results.} 

\subsection{Spaxels} \label{sec_spaxel}

The fitting results we use are from the Pipe3D (\citealt{2016RMxAA..52..171S}; \citealt{2016RMxAA..52...21S}) dataproducts released with SDSS DR14. The Initial Mass Function (IMF) adopted in Pipe3D is \cite{1955ApJ...121..161S}. Assuming the intrinsic flux ratio of $(\ha/\hb)_0$ = 2.86 and a \cite{2000ApJ...533..682C} attenuation law, we then use the Balmer Decrement to correct the emission line fluxes, with which we obtain the SFR, adopting the \cite{1998ARA&A..36..189K} conversion (in Salpeter IMF).

The spaxels are then selected with regard to the following criteria: (1) EW($\ha$) (equivalent width of \ha) $>$ 6 \AA; (2) emission line flux ratio of $\oiii/\hb$ and $\nii/\ha$ lying below the Kewley limit {\citep{2001ApJ...556..121K}} on the BPT diagram {\citep{1981PASP...93....5B}}; (3) signal-to-noise ratio (S/N) of $\ha$ greater than 3 and S/N of other involved emission lines greater than 1. The first two criteria are to select pure star-forming regions, which have been used in many studies ({e.g.} \citealt{2016ApJ...821L..26C}; \citealt{2016A&A...587A..70S}; \citealt{2017arXiv170905438S}). In Pipe3D dataproducts, a spatial binning is performed in order to reach a S/N of 50 measured in the range of 5590--5680 {\AA} across the FoV \citep{2016RMxAA..52..171S}. However, as the binning method requires flux homogeneity in the same bin, some bins would have lower S/Ns than expected. Therefore, we recalculate the continuum S/N for each bin following \cite{2008ASPC..394..505S} and clip those with S/N smaller than 3 {($\sim$4\% of data)}.

\section{Results} \label{sec_results}

\subsection{Resolved Main Sequence for Two Assembly Modes}
\begin{figure*}
\centering
 \includegraphics[width=0.9\textwidth]{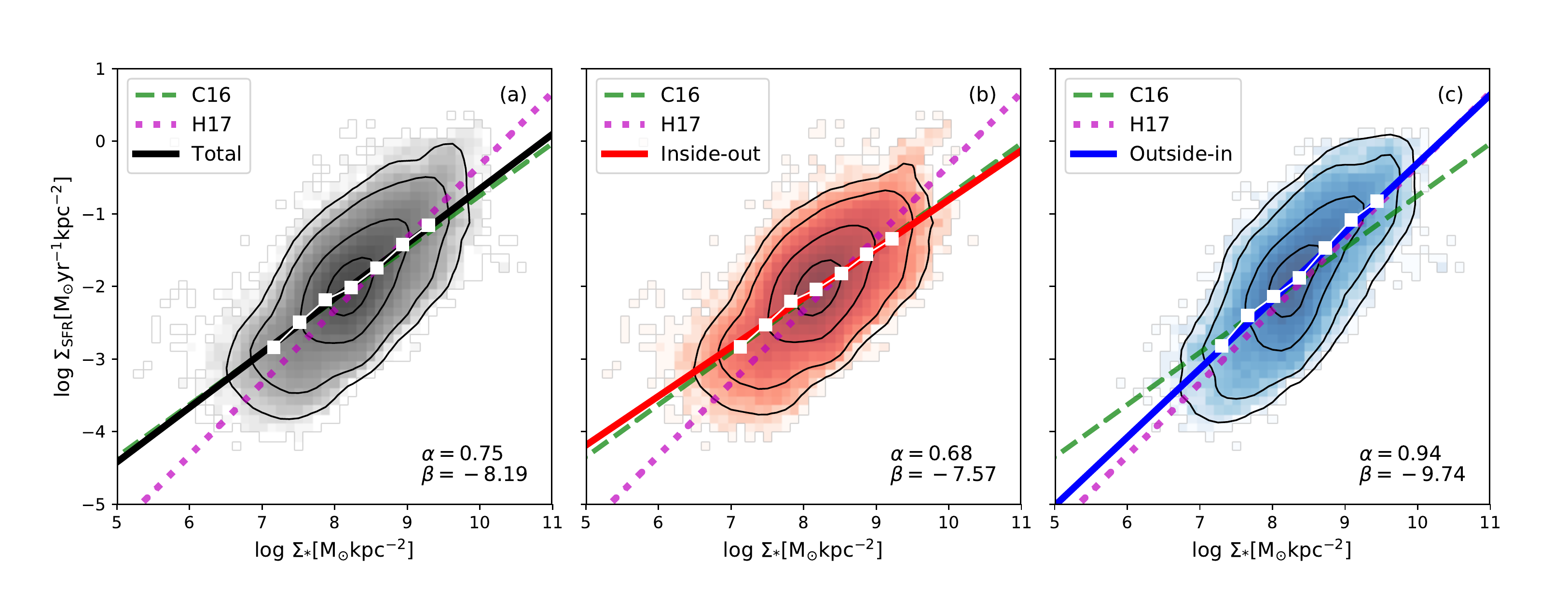}
    \caption{SGMS for our sample and subsamples. White squares show medians of $\Sigma_{\mathrm{SFR}}$ for $\Sigma_{*}$ equally binned between 3\% and 97\% quantiles. The best OLS fittings are shown by colored solid lines in each panel in black (total), red (`inside-out') and blue (`outside-in'). $\alpha$ and $\beta$ are slopes and zero points for the OLS fittings. The OLS fitting for 80\% data of C16 and orthogonal fitting from H17 are shown by {the green dashed line and the magma dotted line, respectively. In each panel, contours show levels of 30\% / 1$\sigma$ (67\%) / 2$\sigma$ (95\%) / 3$\sigma$ (99\%) of data.}}
       \label{Three_SGMS}
\end{figure*} 

In panel (a) of Figure \ref{Three_SGMS}, we show the SGMS for all {141,114} spaxel bins in all the sample galaxies. The black solid line represents the best optimized least square (OLS) linear fitting using all spaxel bins in the form of $\log_{10}(\Sigma_{\mathrm{SFR}})  = \alpha \log_{10}(\Sigma_{*}) + \beta$. The fitted slopes ($\alpha$) and zero points ($\beta$) are shown in each panel. For comparison, results from C16 and H17 are also drawn. With a slope of {0.75}, our result lies between C16 (0.72) and H17 (1.0).

To investigate whether the two subsamples show different patterns on their SGMS, we display the SGMS with regard to the two types of mass assembly modes in panel (b) and (c) of Fig. \ref{Three_SGMS}, respectively. It turns out that the two populations of galaxies occupy different locations on the $\Sigma_{*}$--$\Sigma_{\mathrm{SFR}}$ plane, with `outside-in' galaxies on average lying above `inside-out' galaxies at the high-$\Sigma_{*}$ end (log $(\Sigma_{*})$ $>$ 8.5), indicating an elevation of star formation in these regions in `outside-in' galaxies. Furthermore, the OLS fitting results of the two populations (in red and blue) show that SGMS in `inside-out' galaxies is flatter (with a slope of 0.68) than the total SGMS. In contrast, `outside-in' galaxies exhibit a steeper SGMS (with a slope of {0.94}) than the total one. These indicate that besides distinctions in global properties, they also have clear difference in their resolved star formation patterns.

\subsection{\textit{Galaxy-by-Galaxy} SGMS}

\begin{figure*}
\centering
 \includegraphics[width=0.95\textwidth]{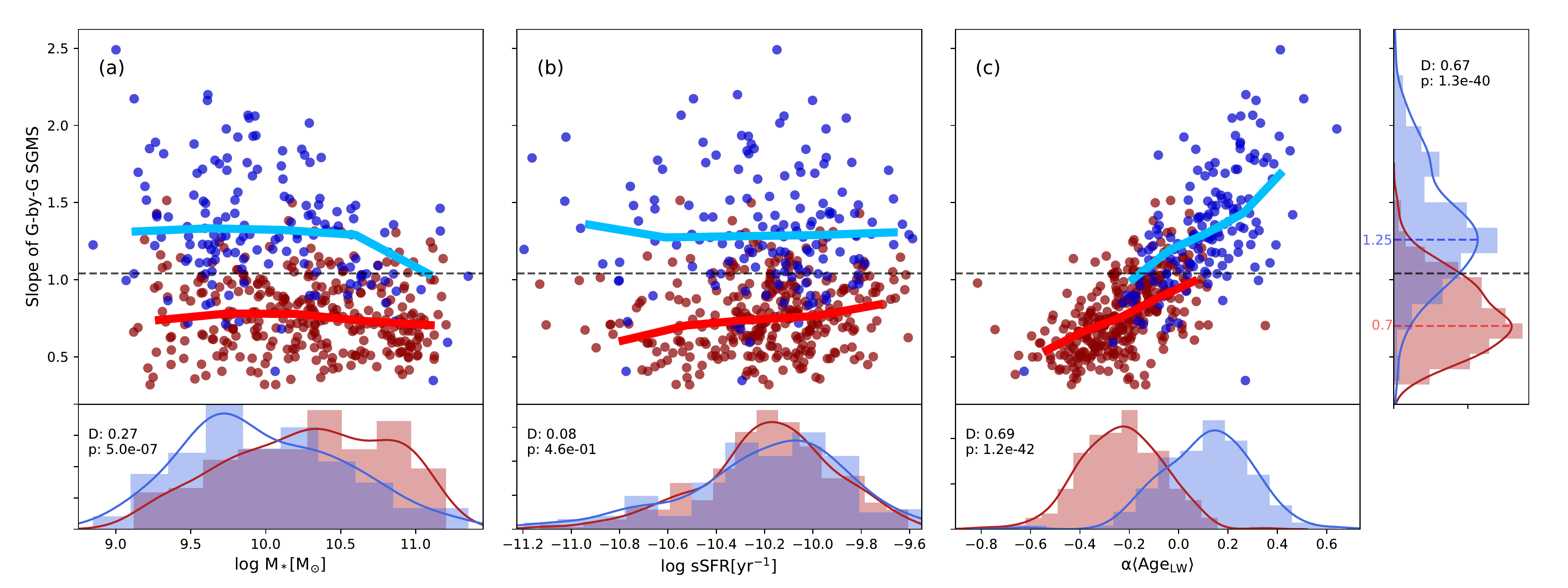}
    \caption{Slopes of G-by-G SGMS with respect to (from left to right) M$_{*}$, sSFR, and $\mathrm{\alpha\langle Age_{LW} \rangle}$. In each panel, `outside-in' and `inside-out' galaxies are represented by blue and red dots. Colored solid lines show trends of five medians of $\Sigma_{\mathrm{SFR}}$ for $\Sigma_{*}$ equally binned between 3\% and 97\% quantiles for the two subsamples. Marginalized distributions of parameters are shown by red (`inside-out' galaxies) and blue (`outside-in' galaxies) histograms in small panels with their results of K--S tests (D-statistics and p-values). Colored dashed lines show their peaks and the black dashed line shows where the SGMS slope equals 1.04.}
       \label{slope-gal-1}
\end{figure*} 

\begin{figure*}
\centering
 \includegraphics[width=0.95\textwidth]{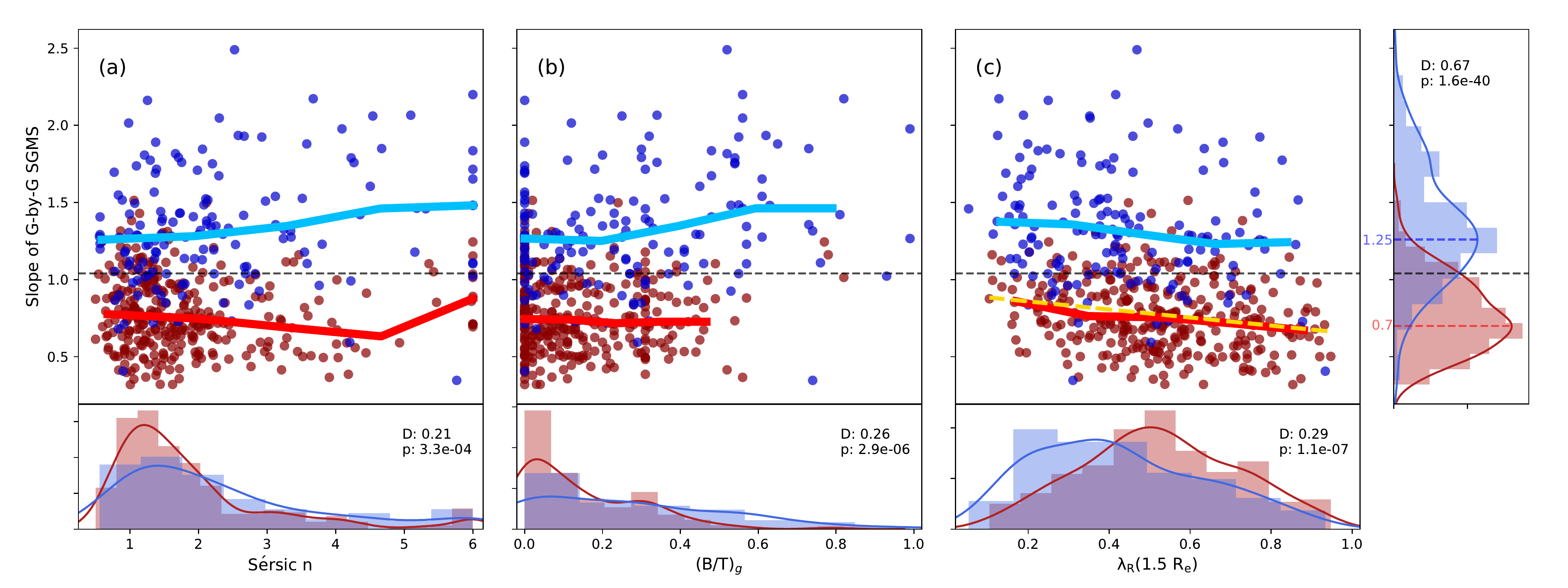}
    \caption{Same plot as Fig. \ref{slope-gal-1} for slopes of G-by-G SGMS with respect to (from left to right) Sersic n, (B/T)$_g$, and $\rm \lambda_R(1.5Re)$. A linear fitting for the red dots is shown by gold dashed lines in panel (c), indicating an anti-correlation between the slope of G-by-G SGMS and $\rm \lambda_R(1.5Re)$ of the galaxy.}
       \label{slope-gal-2}
\end{figure*} 

To explore statistical trends and variations among galaxies, it is also worth inspecting the SGMS within individual galaxies. Therefore, we conduct a similar analysis in \cite{2017MNRAS.466.1192M} by linearly fitting star-forming bins from each galaxy one-by-one. This helps to mitigate the potential problem in resolved studies that the fitting might depend on the sampling since the numbers of bins out of each galaxy are different. For each galaxy, the fitting is iterated for three times to remove outliers beyond 3$\sigma$ regions. We present results for individual galaxies in Figure \ref{slope-gal-1} {and \ref{slope-gal-2}}, excluding the bad fittings with the Pearson correlation coefficient $<$ 0.6 ({$\sim18\%$} of the sample). 

{We perform Kolmogorov--Smirnov (K--S) tests for each marginalized distribution of our two subsamples to test whether they differ, with the D-statistics and p-values shown in Fig. \ref{slope-gal-1} and \ref{slope-gal-2}. If the D-statistic is small or the p-value is high (at a certain significance level), then we cannot reject the hypothesis that their distributions are the same; otherwise, it can be concluded that they are from different distributions.}

In each panel of Fig. \ref{slope-gal-1}, slopes fitted from G-by-G SGMS are plotted as a function of M$_{*}$, specific star formation rate (sSFR) and the luminosity-weighted (LW) age gradient of the host galaxy. 
The distributions of slopes for the two populations are clearly separated. In panel (a) and (b), trends of medians shown in colored solid lines suggest a slight increase for slopes of `inside-out' galaxies with regard to sSFR and weak or no dependence with M$_*$. The dependences are less significant for `outside-in' galaxies as the slopes have a larger scatter. 
{The age gradient is drawn from the Pipe3D catalog in \cite{2017arXiv170905438S}, measured within 0.5--2 R$\rm _e$.}
The well-separated distributions of age gradient in panel (c) are due to the close correlation between D4000 and stellar age (see the sample selection in Section 2.1).

{In each panel of Fig. \ref{slope-gal-2}, we explore the relation of G-by-G SGMS with bulge indicators including the Sersic n index (from the MaNGA DRP catalog), the bulge ratio in G band (B/T)$_g$ (from \citealt{2011ApJS..196...11S}) and the specific angular momentum ($\rm \lambda_R$) of the host galaxy. The $\rm \lambda_R$ parameter is drawn from the Pipe3D catalog \citep{2017arXiv170905438S}, measured within 1.5 R$\rm _e$. This parameter is defined as}
$\rm \lambda_R=\langle R|V|\rangle/ \langle R\sqrt{V^2+\sigma^2}\rangle$
{ (\citealt{2007MNRAS.379..401E}), where V and $\sigma$ are locally measured maximum radial velocity and stellar velocity dispersion, with R representing projected galactocentric distance and brackets symbolizing flux weighting. Panel (a) and Panel (b) show that for `inside-out' galaxies the slopes have weak or no correlations with Sersic n or (B/T)$_g$. On the other hand, `outside-in' galaxies appear to have larger slopes with higher Sersic n or bulge ratios. {However, the Sersic n and B/T parameter may not be directly translated into the morphology because they are not one-to-one tightly correlated (see, e.g. \citealt{2017arXiv170905438S}).} In panel (c), the slopes of `inside-out' galaxies present a slight anti-correlation with $\rm \lambda_R$, i.e. galaxies with higher $\rm \lambda_R(1.5Re)$ have smaller slopes. Their distributions also show that `outside-in' galaxies have systematically smaller $\rm \lambda_R(1.5Re)$, suggesting their higher randomness in stellar motion and kinematic proximity to early-type galaxies \citep{2016ARA&A..54..597C}.}

\subsection{Stacking of G-by-G SGMS}

\begin{figure*}
\centering
 \includegraphics[width=0.95\textwidth]{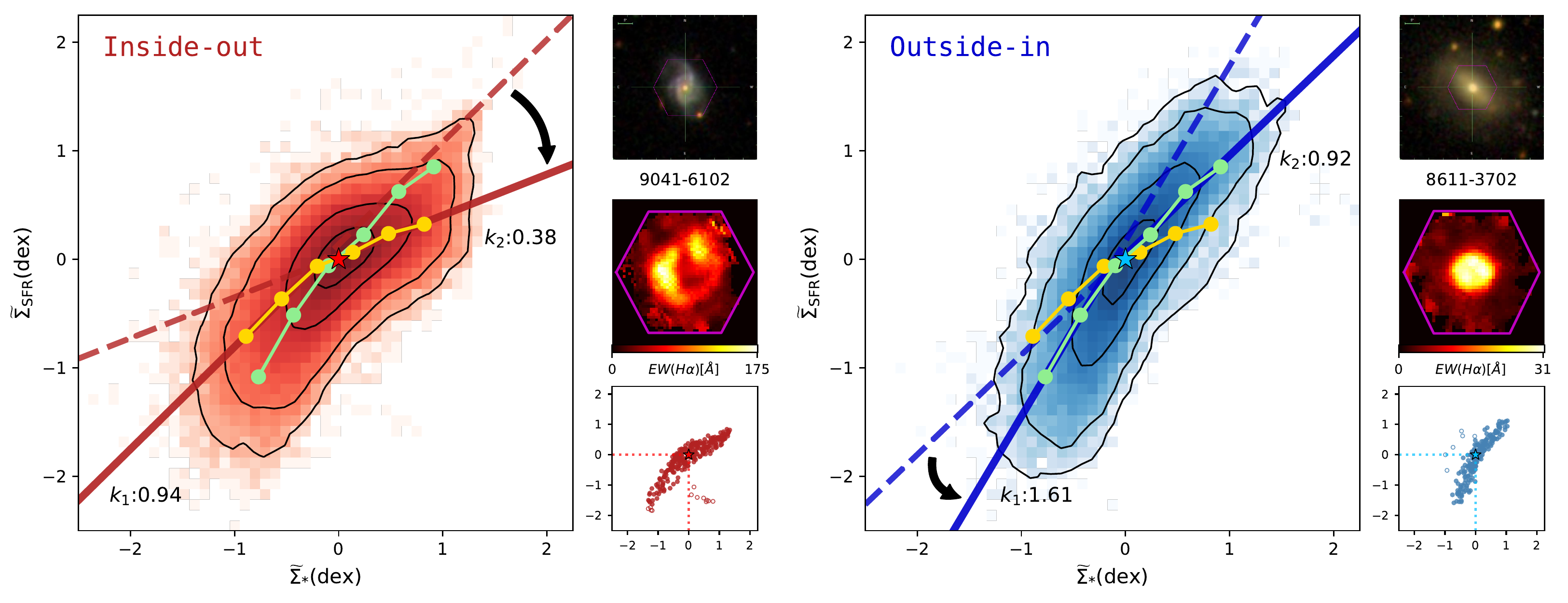}
    \caption{Stacking for G-by-G SGMS of two assembly modes. Colored dots (green for `outside-in' and gold for `inside-out') show modes of $\widetilde{\Sigma}_{\mathrm{SFR}}$ for $\widetilde{\Sigma}_{*}$ in equal bins between 1\% and 99\% quantiles. Two-component linear fittings for the modes are show by solid lines, with dashed lines representing their extrapolations. The inner parts of `inside-out' and outer parts of `outside-in' galaxies appear to be in good alignment with slopes close to 0.9, which is consistent with results of \cite{2016MNRAS.456.4533M} and \cite{2017MNRAS.466.1192M}. Contours show levels of 30\% / 1$\sigma$ (67\%) / 2$\sigma$ (95\%) / 3$\sigma$ (99\%) of data. {SDSS gri images, EW($\ha$) maps and G-by-G SGMS of two typical galaxies (MaNGA ID: 9041-6102 and 8611-3702) showing a clear piecewise pattern on a G-by-G level are shown in small panels for each subsample. In the bottom small panels, outliers $>$ 3$\sigma$ are shown by empty circles. Colored stars show the characteristic values of SGMS, which are shifted to zero.}}
       \label{Stacking}
 \end{figure*}
 
The fitting of G-by-G SGMS also shows differences in the zero points among galaxies, which might be attributed to differences in their physical properties or conditions. For this reason, we shift the SGMS of each individual galaxy to have the median values of $\Sigma_{*}$ and $\Sigma_{\mathrm{SFR}}$ equal to zero, and stack them on one panel for `inside-out' and  `outside-in' galaxies to see their statistical patterns. This is equivalent to select the characteristic values of $\Sigma_{*}$ and $\Sigma_{\mathrm{SFR}}$ for each galaxy and apply a normalization on them. The results of stacking SGMS for the two subsamples are shown in Figure \ref{Stacking}. The shifted $\Sigma_{*}$ and $\Sigma_{\mathrm{SFR}}$ are referred to as $\widetilde{\Sigma}_{*}$ and $\widetilde{\Sigma}_{\mathrm{SFR}}$.

Firstly, the small scatter ($\sim$0.3 dex) implies the tightness of the spatially resolved correlation. Secondly, a quick visual inspection reveals a conspicuous flattening at the high-$\Sigma_{*}$ end for `inside-out' galaxies and a less significant but perceptible steepening at the low-$\Sigma_{*}$ end for `outside-in' galaxies. Such patterns are hard to observe in Fig. \ref{Three_SGMS} since differences in zero points mix the SGMS together. To further illustrate, we apply two-component linear fittings on modes of $\Sigma_{\mathrm{SFR}}$ for $\Sigma_{*}$ equally binned between 1\% and 99\% quantiles, with a restriction that the turning points lie within the central 40\% of data. A cross-validation technique \citep{2014sdmm.book.....I} is applied to demonstrate that the piecewise fittings have less variance, which also have smaller $\chi^2_{dof}$ when uncertainties of $\widetilde{\Sigma}_{\mathrm{SFR}}$ are considered. However, we emphasize that the main purpose of fitting is to strengthen the visual inspection. 

The slopes of SGMS in the literature are dependent on the sample selection, analyzing method, and fitting recipe; however, they all range from {0.66} to 1.0 ({e.g. \citealt{2013A&A...554A..58S}}; \citealt{2013ApJ...779..135W}; \citealt{2016ApJ...821L..26C}; \citealt{2016MNRAS.456.4533M}; \citealt{2017MNRAS.469.2806A}; \citealt{2017ApJ...851L..24H}; \citealt{2017MNRAS.466.1192M}). 
In Fig. \ref{Stacking}, the fitting slopes for outer parts of `inside-out' galaxies and inner parts of `outside-in' galaxies are close to 0.9, which is situated within the range given in previous studies and is well consistent with results of \cite{2016MNRAS.456.4533M} and \cite{2017MNRAS.466.1192M}. {Thus, we treat them as the standard pattern of star formation.} The remaining parts show a deviation from them, especially in `inside-out' galaxies. This suggests that the expected in-situ star formation in the outer regions of `inside-out' galaxies and the inner regions of `outside-in' galaxies are governed by a common set of regulations without the influence of additional physical processes, and thus we infer that the inner parts of `inside-out' galaxies clearly show an evidence of suppression in star formation. For `outside-in' galaxies, they also show an indication of suppression in star formation for their outer regions, as they appear to have insufficient SFR at certain surface density (which leads to a steeper slope). {We have stacked the two panels of Fig. \ref{Stacking} into one and find that the outer regions of `outside-in' galaxies indeed have different distributions from `inside-out' galaxies. We will have more discussions on this in the next section. We further perform a test with a lower continuum S/N criteria to confirm that this pattern is not due to an S/N cut in lower surface density regions.}
 
\section{DISCUSSION} \label{sec_discussion}

As confirmed by many previous studies, the SGMS holds well down to kiloparsec {scales}. In this work, our results further reveal that, the recent mass assembly mode of the galaxy would have impacts on the shape of their SGMS. It is worth noting that our subsample construction for `inside-out' and `outside-in' galaxies is different from the fossil record method used in many previous studies ({e.g.} \citealt{2013ApJ...764L...1P}; \citealt{2016MNRAS.463.2799I}; \citealt{2017A&A...608A..27G}). Given the short timescales and stochasticity for star-forming activities, the mass assembly mode of the galaxy is possibly not fixed.  For the same reason, it is reasonable to infer that the SGMS is more related to the recent mass assembly mode, which can be well traced by the D4000 diagnostic. Furthermore, it is model-independent with fewer biases, since different fitting methods might lead to different results \citep{2017MNRAS.466.4731G}. {However, we notice here the potential caveat of using D4000 as the indicator is that, although primarily correlated with the age of the stellar population, D4000 also depends on the metallicity (\citealt{1997A&A...325.1025P}), which is related to the age-color degeneracy.} To check the mass assembly histories of our sample, we have also stacked and compared the mass growth curves in inner and outer regions. We find that $>$ 40\% of galaxies have complicated overlapping mass growth curves hard to define their assembly histories, which supports our previous speculation and is consistent with episodic transitions proposed in \cite{2016MNRAS.463.2799I}.

\subsection{`Inside-out' SGMS: Bulge Effects or Central Suppression?}

The smaller slope of `inside-out' SGMS mainly comes from the suppression in their central high surface density parts, which has been observed in many recent studies ({e.g.} \citealt{2017MNRAS.469.2806A}; \citealt{2018MNRAS.474.2039E}; \citealt{2018MNRAS.476..580S}). This can be explained {by the inside-out star formation cessation as proposed in the literature (e.g. \citealt{2015Sci...348..314T}, \citealt{2017arXiv170905438S}, \citealt{2017arXiv171007569W}, \citealt{2018MNRAS.tmp..742B}), probably accompanied with deficiencies in gas} either caused by stellar feedback or AGN feedback. G-by-G results {appear to be in agreement with} this, with some even having negative slopes around their centers.

{An alternative interpretation holds considering that the ongoing star formation is more relevant to the disk where cold gas condenses and forms stars. In this case, the apparent bend can be explained by the presence of the classical bulge that is less correlated with the galaxy self-regulation, in which the deficiency in cold gas and a `physical' suppression is not necessary. However, as shown in Fig. \ref{slope-gal-2}, `outside-in' galaxies have a broader distribution in (B/T)$_g$ and systematically larger $\rm \lambda_R$, indicating their relative significance in presence of the bulge from both photometric and kinematic perspectives. Meanwhile, they do not present a significant bend in their centers. Therefore, we infer that, at least in our sample, the presence of the bulge only is not a predominant factor contributing to the bend of SGMS.}

\begin{figure}
 \includegraphics[width=8.5cm]{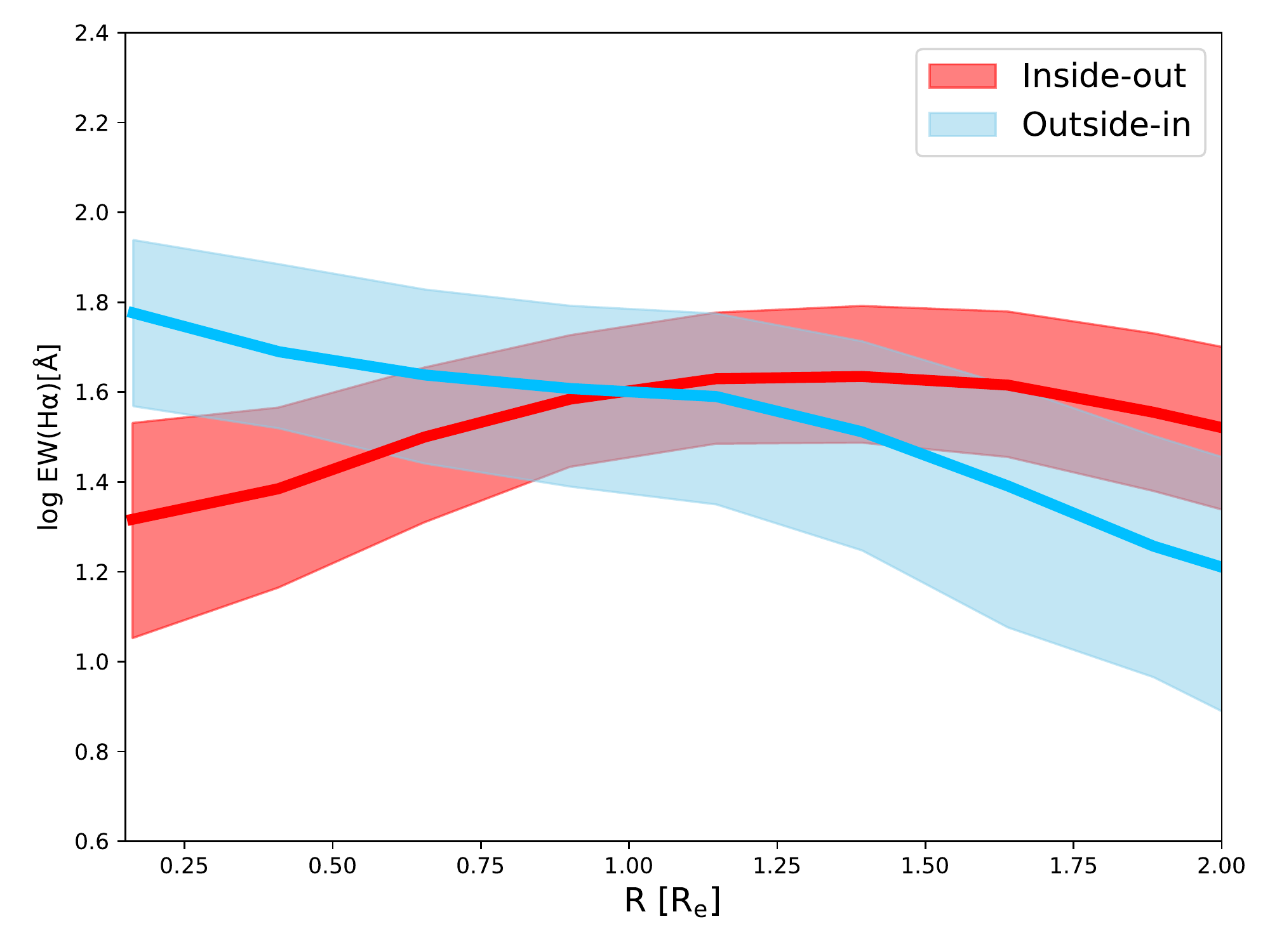}
    \caption{{EW(\ha) profiles of `inside-out' galaxies (red) and `outside-in' galaxies (blue) {equally sampled along the radius}. The solid lines stand for the median profiles of EW(\ha) and the translucent bands show the {30\%--70\%} distributions. Star formation in `inside-out' galaxies peaks in disks, wile star formation in `outside-in' galaxies peaks in central regions and decreases with the radius.}}
       \label{EW_Ha_profile}
 \end{figure}

{In Figure \ref{EW_Ha_profile}, we plot profiles of EW(\ha) for `inside-out' galaxies and `outside-in' galaxies in red and blue, respectively, where it can be seen that `inside-out' galaxies show a strong decrease with their star formation peaking in disks. Recently, \cite{2018MNRAS.tmp..742B} have demonstrated that EW(\ha) and sSFR profiles of SFGs with higher M$_*$, especially for those with M$_*$ $>$ 10$^{10.5}$, generally present a stronger decrease at their centers. \cite{2016A&A...590A..44G} have observed a similar trend with the morphology, since morphologies of galaxies correlate with M$_*$. Our result is consistent with \cite{2018MNRAS.tmp..742B}, where we check that more massive galaxies have lower EW(\ha) in their centers. It could be inferred that the difference of profiles observed in Fig. \ref{EW_Ha_profile} is due to the fact that `inside-out' galaxies are systematically more massive than `outside-in' galaxies. However, by constructing control samples for `inside-out' and  `outside-in' galaxies in M$_*$/$\rm \lambda_R$ with no significant changes observed, i.e. their overall trends remain, we confirm that the difference of profiles of our subsamples in inner regions is not led by their mass/morphology distinctions.}

\subsection{`Outside-in' SGMS: Standard or Peripheral Suppression?}

On the other hand, the larger slope of the `outside-in' SGMS approaching (in the pixel-by-pixel case) or even exceeding (in G-by-G cases) unity is worth investigation. {When compared with `inside-out' galaxies, one would naturally infer that the `outside-in' galaxies present a `standard' star-forming pattern in both inner regions and outskirts. This point might be supported by the similarity in its slope to the high-redshift integrated SFMS (e.g. \citealt{2014ApJS..214...15S}, \citealt{2017ApJ...840...47B}), since galaxies in that epoch contain a higher fraction of star-forming regions. However, many `outside-in' galaxies actually present a `truncated' star-forming pattern, which can be revealed in Fig. \ref{EW_Ha_profile} that `outside-in' galaxies typically {present} a decreasing trend in star formation along radius {(and decreases faster after reaching 1.25 R$\rm _e$)} instead of a flat profile. Therefore, we consider that they are not forming stars in a uniform manner throughout their optical extensions.} 

Our result for `outside-in' galaxies is in good agreement with the compaction model, which predicts the central enhanced star formation (\citealt{2014MNRAS.438.1870D}; \citealt{2016MNRAS.457.2790T}; \citealt{2016MNRAS.458..242T}) accompanied by the suppression or consistency \citep{2016MNRAS.458..242T} in outer regions. This is what we have observed in the case of stacking SGMS for `outside-in' galaxies (Fig. \ref{Stacking}) and in some of their G-by-G SGMS (but not all). These galaxies have higher surface densities and are more concentrated in optical morphologies \citep{2017ApJ...844..144W}. We also have checked that star formation in `outside-in' galaxies tend to be more compact and centralized, by measuring the concentration parameter \citep{2014ARA&A..52..291C} of their EW($\ha$) maps, which are roughly equivalent to sSFR maps. The preferred locations above the main sequence for `outside-in' galaxies with M$_*$ $>$ 10$^{10}$ M$_{\odot}$ are also consistent with \cite{2016MNRAS.457.2790T} and correspond to the central enhanced SFGs observed in \cite{2018MNRAS.474.2039E}. The compaction process may be attributed to a series of physical mechanisms, such as disk instability, bar-induced inflow, and interaction with the environment. {This process is expected to happen more frequently in high-redshift cosmic epochs (z $>$ 1), but qualitatively match with some of our `outside-in' subsample in many aspects.} For some of the `outside-in' galaxies in our subsample, especially those relatively massive ones, they might have accreted cold gas and transported them to central regions for star formation through gravitational disk instability or the presence of bars{, and thus appear to be in the compaction process}. 
 
{However, we also note that the compaction model mainly operates in the range of massive galaxies and thus cannot explain the whole population. Therefore, alternatively}, environmental quenching effects \citep{2014A&ARv..22...74B} might also play a role, {especially for those less massive ones,} given their higher over-density environments on average \citep{2017ApJ...844..144W}. The outskirts of the galaxies could have their gas reservoirs removed through effects such as gas stripping (\citealt{1972ApJ...176....1G}; \citealt{1999MNRAS.308..947A}). As a consequence, the intensity of star formation in the galaxy outskirt is lower than expected. Meanwhile, disk instability and gas inflow caused by tidal forces could trigger and then support the central star formation. Whether spontaneously or passively, `outside-in' galaxies are expected to quench, which is consistent with the conclusion in \cite{2017ApJ...844..144W}. Although the current picture is still elusive, future information from the distribution of the gas content would shed more light on their natures.

\section{CONCLUSION} \label{sec_conclusion}

We investigate the spatially resolved SFMS in $\sim$600 {face-on} SFGs using IFS data from the MaNGA survey. The quantities we use are from SDSS-MaNGA-Pipe3D dataproducts. We find that the SGMS holds well down to the {1--2 kpc scale for our sample}. Using resolved information from the 4000 {\AA} break, we divide our sample into two subsamples, according to their recent mass assembly modes. We find that they show distinct behaviors on the $\Sigma_{*}$--$\Sigma_{\mathrm{SFR}}$ plane, with `outside-in' galaxies appearing to be steeper, and more elevated than `inside-out' galaxies at higher surface densities. We further inspect the SGMS in individual galaxies and find their distributions of slopes to be clearly separated. `Inside-out' galaxies show an increase in their slopes with increasing sSFR {and an anti-correlation with the specific angular momentum}, while `outside-in' galaxies have a larger scatter in their distribution of slopes. By normalizing and stacking the SGMS for each galaxy, we find a clear suppression in inner regions of `inside-out' galaxies and a less significant suppression in outer regions of `outside-in' galaxies, leaving the remaining parts following a slightly sublinear scaling relation with a slope of $\sim0.9$. We attribute the difference of behaviors on the SGMS for the two populations to the differences in their evolutionary stages or environments.

\section*{ACKNOWLEDGEMENT}

This work is supported by the National Key R\&D Program of China (2015CB857004, 2017YFA0402600) and the National Natural Science Foundation of China (NSFC, Nos. 11320101002, 11421303, and 11433005). E.W. was supported by the Youth Innovation Fund by the University of Science and Technology of China (No. WK2030220019). Q.L. was supported by Fund for Fostering Talents in Basic Science of the National Natural Science Foundation of China NO.J1310021. Q.L. sincerely thanks Lin Lin, Guilin Liu, Lei Hao, Yicheng Guo, Sandra Faber, David Koo and Hassen Yesuf for helpful discussions. The authors acknowledge the referee for her/his constructive and useful comments which improved our original manuscript.\smallskip

Funding for the Sloan Digital Sky Survey IV has been provided by
the Alfred P. Sloan Foundation, the U.S. Department of Energy Office of
Science, and the Participating Institutions. SDSS-IV acknowledges
support and resources from the Center for High-Performance Computing at
the University of Utah. The SDSS website is www.sdss.org.
SDSS-IV is managed by the Astrophysical Research Consortium for the
Participating Institutions of the SDSS Collaboration including the
Brazilian Participation Group, the Carnegie Institution for Science,
Carnegie Mellon University, the Chilean Participation Group, the French Participation Group, 
Harvard-Smithsonian Center for Astrophysics,
Instituto de Astrof\'isica de Canarias, The Johns Hopkins University,
Kavli Institute for the Physics and Mathematics of the Universe (IPMU) /
University of Tokyo, Lawrence Berkeley National Laboratory,
Leibniz Institut f\"ur Astrophysik Potsdam (AIP),
Max-Planck-Institut f\"ur Astronomie (MPIA Heidelberg),
Max-Planck-Institut f\"ur Astrophysik (MPA Garching),
Max-Planck-Institut f\"ur Extraterrestrische Physik (MPE),
National Astronomical Observatory of China, New Mexico State University,
New York University, University of Notre Dame,
Observat\'ario Nacional / MCTI, The Ohio State University,
Pennsylvania State University, Shanghai Astronomical Observatory,
United Kingdom Participation Group,
Universidad Nacional Aut\'onoma de M\'exico, University of Arizona,
University of Colorado Boulder, University of Oxford, University of Portsmouth,
University of Utah, University of Virginia, University of Washington, University of Wisconsin,
Vanderbilt University, and Yale University.

This project makes use of the MaNGA-Pipe3D dataproducts. We thank the IA-UNAM MaNGA team for creating this catalog, and the ConaCyt-180125 project for supporting them.


\appendix 
\counterwithin{figure}{section}
\section{Resolution Effect with MaNGA FWHM PSF} \label{sec:appendix}

{The reconstructed PSF of the MaNGA datacube is 2.5$\arcsec$ in FWHM. Therefore, according to the Nyquist sampling theorem, those galaxies in the MaNGA sample with R$\rm _e<$2.5$\arcsec$(1 PSF) are not resolved, and those with 2.5$\arcsec$ $<$ R$\rm _e<$ 5$\arcsec$ (1$\sim$2 PSF) might be affected by resolution effects that spatially resolved quantities would be flattened, thereby changing the shape of SGMS. In fact, \cite{2014A&A...561A.129M} concludes that one should take careful considerations of effects from the resolution and redshift distribution of the sample, when using data from IFU surveys such as MaNGA to investigate spatially resolved properties. Therefore, it is crucial to demonstrate that resolution effects would not significantly affect our results.}

\begin{figure}[htp]
 \includegraphics[width=7.cm]{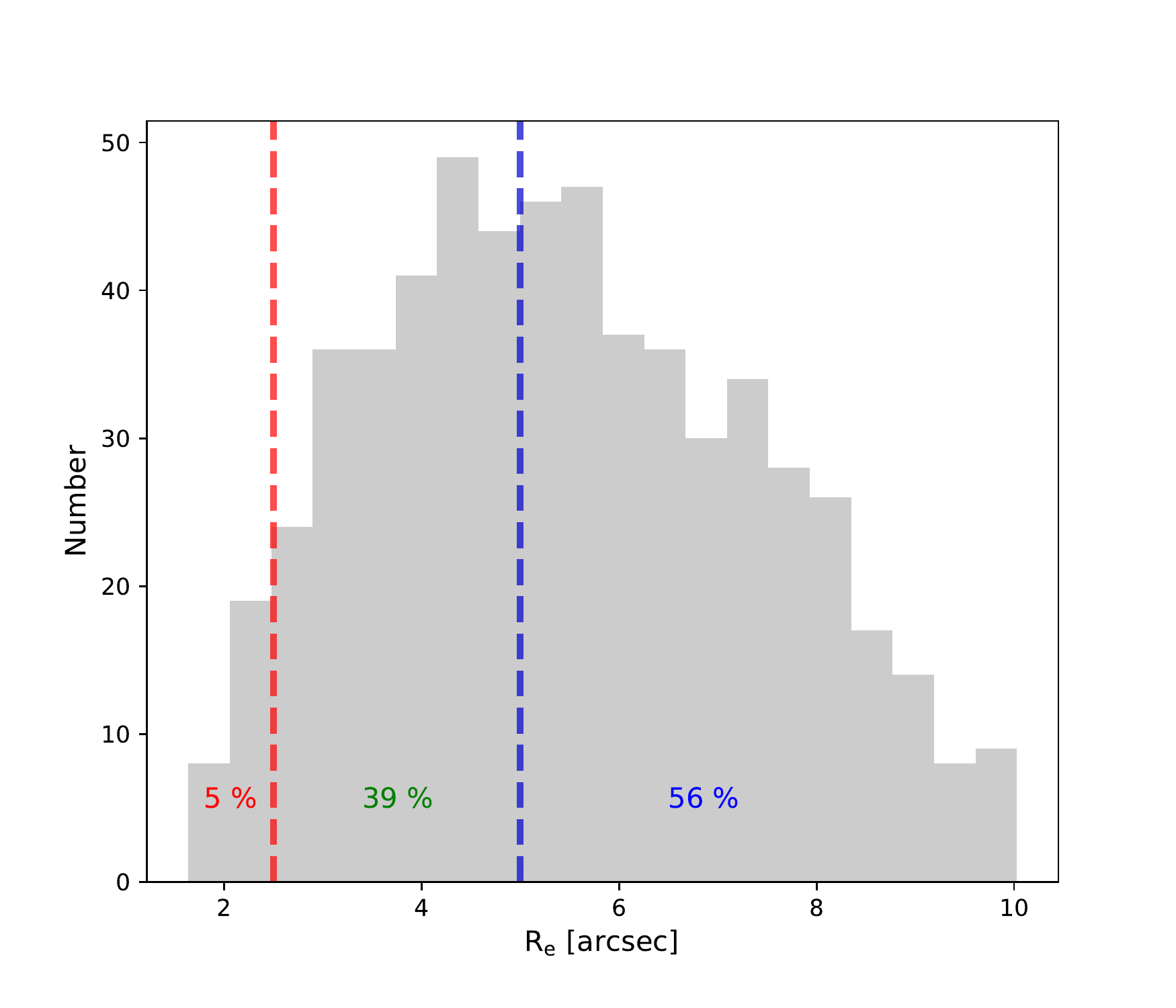}
 \centering
    \caption{R$\rm _e$ histogram for our 561 MaNGA galaxies used in the main analysis.  The red dashed line stands for the MaNGA FWHM PSF (2.5$\arcsec$), and the blue dashed line stands for two times of PSF (5$\arcsec$). Fractions of galaxies lying between the PSF limits are indicated in numbers at the bottom.}
       \label{Re_histogram}
 \end{figure}

\begin{figure}[htp]
 \includegraphics[width=0.75\textwidth]{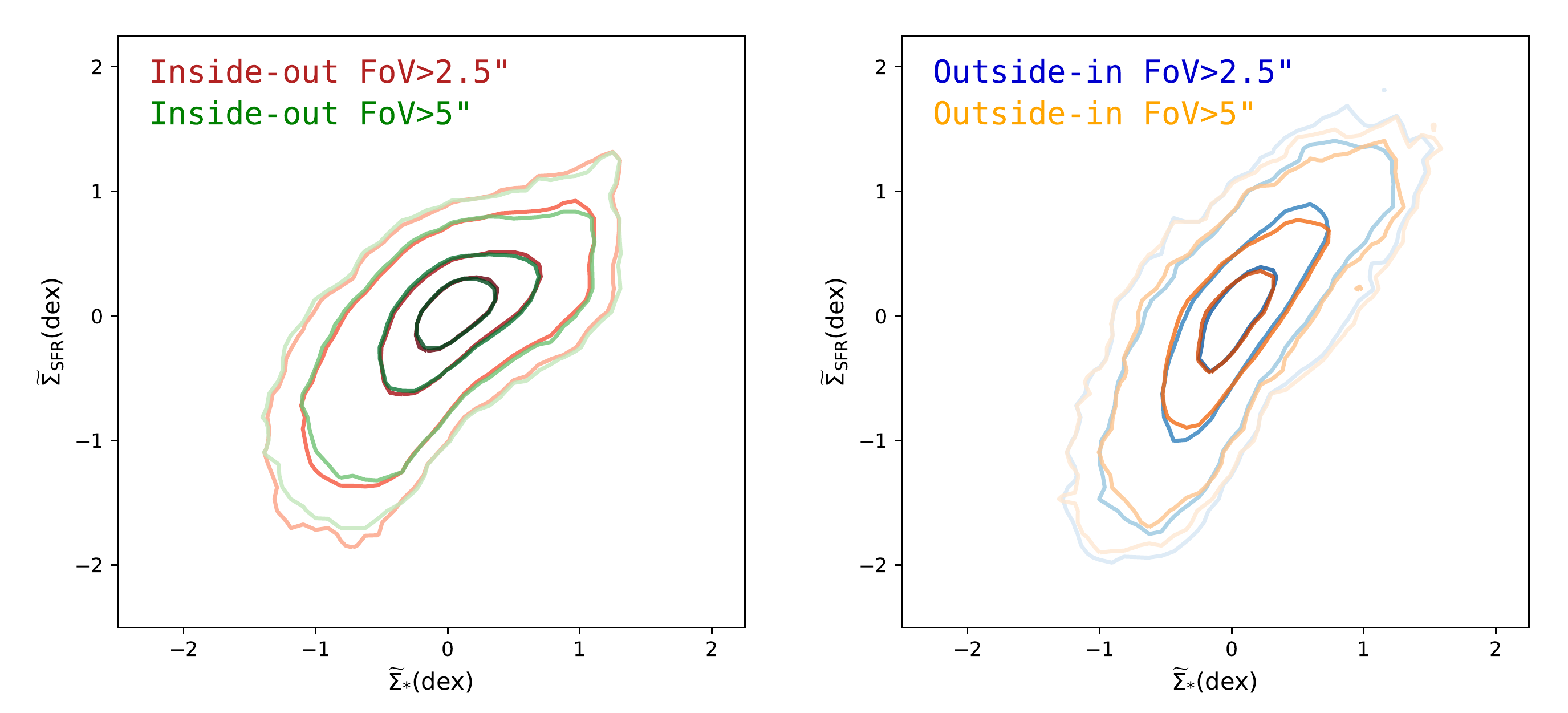}
 \centering
    \caption{{Contours showing 30\% / 1$\sigma$ (67\%) / 2$\sigma$ (95\%) / 3$\sigma$ (99\%) distributions of the stacking of G-by-G SGMS before and after excluding galaxies with PSF $<$ R$\rm _e<2$ PSF. Left: for `inside-out' galaxies, red lines represent galaxies with R$\rm _e<$ PSF whereas green lines represent those with R$\rm _e>2$ PSF. Right: for `outside-in' galaxies, blue lines represent galaxies with R$\rm _e<$ PSF whereas orange lines represent those with R$\rm _e>2$ PSF.}}
       \label{Stacking_2PSF}
 \end{figure}

\begin{figure}[htp]
 \includegraphics[width=7.cm]{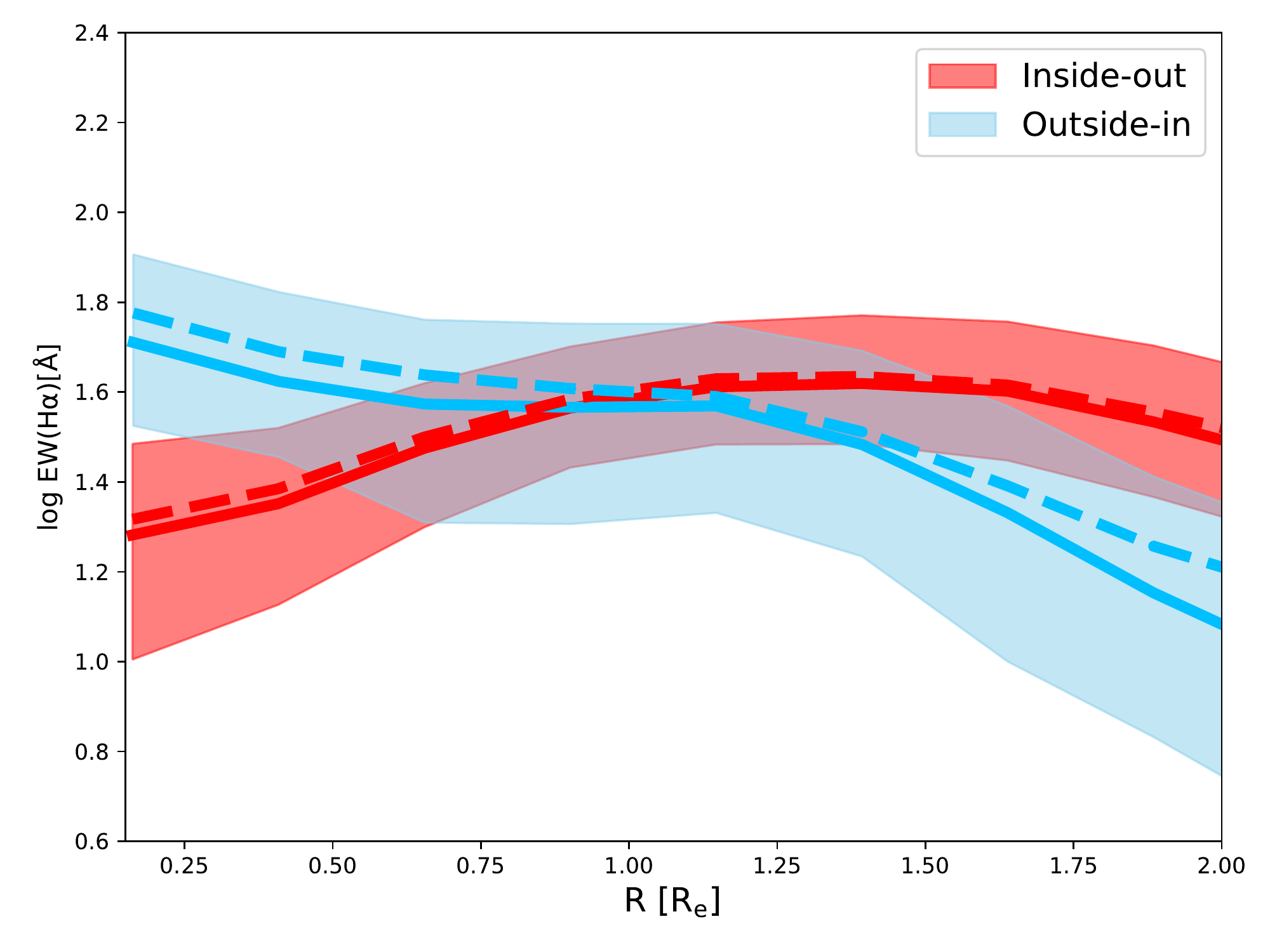}
 \centering
 \caption{{Reproduction of Fig. \ref{EW_Ha_profile} using galaxies with R$\rm _e>2$ PSF (56\% of data). Solid lines show the median profiles of EW(\ha) in red (`inside-out' galaxies) and blue (`outside-in' galaxies), whereas the bands represent their 30\%--70\% distributions. For comparison, profiles using galaxies with R$\rm _e$ $>$ PSF (95\% of data) are shown in dashed lines (i.e. solid lines in Fig. \ref{EW_Ha_profile}).}}
       \label{EW_Ha_profile_2PSF}
\end{figure}

{In Figure \ref{Re_histogram}, we show the R$\rm _e$ distribution of our whole sample, including both `inside-out' and `outside-in' galaxies. The red dashed line represents the FWHM PSF of MaNGA and the blue dashed line represents two times of the PSF of MaNGA. For our sample, about 5\% of the whole population have R$\rm _e$ smaller than the FWHM, which have already been excluded in our main analysis because these galaxies might induce severe bias. About 39\% of our sample have R$\rm _e$ lying between 1$\sim$2 PSF, which are not perfectly resolved. Considering that an exclusion of these galaxies would significantly reduce our sample size, we still include them in our analysis. Therefore, we conduct a test by removing these galaxies to inspect whether the results change or not, which is shown in Figure \ref{Stacking_2PSF} and Figure \ref{EW_Ha_profile_2PSF}.}
 
{Fig. \ref{Stacking_2PSF} shows the distributions of our stacking result of G-by-G SGMS in contours, with solid lines in red/blue standing for the `inside-out'/`outside-in' sample used in the main text and in green/orange for those with R$\rm _e>2$ PSF. Fig. \ref{EW_Ha_profile_2PSF} shows the reproduction of Fig. \ref{EW_Ha_profile}, where profiles of `inside-out'/`outside-in' galaxies with R$\rm _e>2$ PSF are represented by red/blue solid lines, with the profiles in the main text are represented by dashed lines for comparison. For both figures, no significant changes are observed, except for a slight bend for `outside-in' galaxies. This is mainly because `outside-in' galaxies mostly appear to be compact in their optical morphology \citep{2017ApJ...844..144W} with high concentration parameters and small measured values of R$\rm _e$, and thus they are more susceptible to resolution effects. However, the overall distributions and trends remain the same; therefore, we conclude that our results are not significantly influenced by resolution effects and our main conclusions still hold.}

\newpage

\begin{thebibliography}{}

\bibitem[Abadi et al.(1999)]{1999MNRAS.308..947A} Abadi, M.~G., Moore, B., \& Bower, R.~G.\ 1999, \mnras, 308, 947 


\bibitem[Abdurro'uf \& Akiyama(2017)]{2017MNRAS.469.2806A} Abdurro'uf, \& Akiyama, M.\ 2017, \mnras, 469, 2806 


\bibitem[Abolfathi et al.(2017)]{2017arXiv170709322A} Abolfathi, B., Aguado, D.~S., Aguilar, G., et al.\ 2017, arXiv:1707.09322 


\bibitem[Bacon et al.(2017)]{2017A&A...608A...1B} Bacon, R., Conseil, S., Mary, D., et al.\ 2017, \aap, 608, A1 


\bibitem[Baldwin et al.(1981)]{1981PASP...93....5B} Baldwin, J.~A., Phillips, M.~M., \& Terlevich, R.\ 1981, \pasp, 93, 5 


\bibitem[Balogh et al.(1999)]{1999ApJ...527...54B} Balogh, M.~L., Morris, S.~L., Yee, H.~K.~C., Carlberg, R.~G., \& Ellingson, E.\ 1999, \apj, 527, 54 


\bibitem[Barro et al.(2017)]{2017ApJ...840...47B} Barro, G., Faber, S.~M., Koo, D.~C., et al.\ 2017, \apj, 840, 47 


\bibitem[Belfiore et al.(2018)]{2018MNRAS.tmp..742B} Belfiore, F., Maiolino, R., Bundy, K., et al.\ 2018, \mnras,


\bibitem[Boselli \& Gavazzi(2014)]{2014A&ARv..22...74B} Boselli, A., \& Gavazzi, G.\ 2014, \aapr, 22, 74 


\bibitem[Brinchmann et al.(2004)]{2004MNRAS.351.1151B} Brinchmann, J., Charlot, S., White, S.~D.~M., et al.\ 2004, \mnras, 351, 1151 


\bibitem[Bruzual A.(1983)]{1983ApJ...273..105B} Bruzual A., G.\ 1983, \apj, 273, 105 


\bibitem[Bundy et al.(2015)]{2015ApJ...798....7B} Bundy, K., Bershady, M.~A., Law, D.~R., et al.\ 2015, \apj, 798, 7 


\bibitem[Calzetti et al.(2000)]{2000ApJ...533..682C} Calzetti, D., Armus, L., Bohlin, R.~C., et al.\ 2000, \apj, 533, 682 


\bibitem[Cano-D{\'{\i}}az et al.(2016)]{2016ApJ...821L..26C} Cano-D{\'{\i}}az, M., S{\'a}nchez, S.~F., Zibetti, S., et al.\ 2016, \apjl, 821, L26 


\bibitem[Cappellari(2016)]{2016ARA&A..54..597C} Cappellari, M.\ 2016, \araa, 54, 597 


\bibitem[Conselice(2014)]{2014ARA&A..52..291C} Conselice, C.~J.\ 2014, \araa, 52, 291 


\bibitem[Daddi et al.(2007)]{2007ApJ...670..156D} Daddi, E., Dickinson, M., Morrison, G., et al.\ 2007, \apj, 670, 156 


\bibitem[Dekel \& Burkert(2014)]{2014MNRAS.438.1870D} Dekel, A., \& Burkert, A.\ 2014, \mnras, 438, 1870 


\bibitem[Elbaz et al.(2007)]{2007A&A...468...33E} Elbaz, D., Daddi, E., Le Borgne, D., et al.\ 2007, \aap, 468, 33 


\bibitem[Ellison et al.(2018)]{2018MNRAS.474.2039E} Ellison, S.~L., S{\'a}nchez, S.~F., Ibarra-Medel, H., et al.\ 2018, \mnras, 474, 2039


\bibitem[Emsellem et al.(2007)]{2007MNRAS.379..401E} Emsellem, E., Cappellari, M., Krajnovi{\'c}, D., et al.\ 2007, \mnras, 379, 401 


\bibitem[Gallazzi et al.(2005)]{2005MNRAS.362...41G} Gallazzi, A., Charlot, S., Brinchmann, J., White, S.~D.~M., \& Tremonti, C.~A.\ 2005, \mnras, 362, 41


\bibitem[Garc{\'{\i}}a-Benito et al.(2017)]{2017arXiv170900413G} Garc{\'{\i}}a-Benito, R., Gonz{\'a}lez Delgado, R.~M., P{\'e}rez, E., et al.\ 2017, arXiv:1709.00413 

\bibitem[Garc{\'{\i}}a-Benito et al.(2017)]{2017A&A...608A..27G} Garc{\'{\i}}a-Benito, R., Gonz{\'a}lez Delgado, R.~M., P{\'e}rez, E., et al.\ 2017, \aap, 608, A27 


\bibitem[Goddard et al.(2017)]{2017MNRAS.466.4731G} Goddard, D., Thomas, D., Maraston, C., et al.\ 2017, \mnras, 466, 4731 


\bibitem[Gonz{\'a}lez Delgado et al.(2016)]{2016A&A...590A..44G} Gonz{\'a}lez Delgado, R.~M., Cid Fernandes, R., P{\'e}rez, E., et al.\ 2016, \aap, 590, A44 


\bibitem[Gunn \& Gott(1972)]{1972ApJ...176....1G} Gunn, J.~E., \& Gott, J.~R., III 1972, \apj, 176, 1 


\bibitem[Hsieh et al.(2017)]{2017ApJ...851L..24H} Hsieh, B.~C., Lin, L., Lin, J.~H., et al.\ 2017, \apjl, 851, L24


\bibitem[Ibarra-Medel et al.(2016)]{2016MNRAS.463.2799I} Ibarra-Medel, H.~J., S{\'a}nchez, S.~F., Avila-Reese, V., et al.\ 2016, \mnras, 463, 2799 


\bibitem[Ivezi{\'c} et al.(2014)]{2014sdmm.book.....I} Ivezi{\'c}, {\v Z}., Connelly, A.~J., VanderPlas, J.~T., \& Gray, A.\ 2014, Statistics, Data Mining, and Machine Learningin Astronomy, by Z.~Ivenci{\'c} et al.~Princeton, NJ: Princeton University Press, 2014,



\bibitem[Kauffmann et al.(2003)]{2003MNRAS.341...33K} Kauffmann, G., Heckman, T.~M., White, S.~D.~M., et al.\ 2003, \mnras, 341, 33 


\bibitem[Kennicutt(1998)]{1998ARA&A..36..189K} Kennicutt, R.~C., Jr.\ 1998, \araa, 36, 189 


\bibitem[Kewley et al.(2001)]{2001ApJ...556..121K} Kewley, L.~J., Dopita, M.~A., Sutherland, R.~S., Heisler, C.~A., \& Trevena, J.\ 2001, \apj, 556, 121 


\bibitem[Li et al.(2015)]{2015ApJ...804..125L} Li, C., Wang, E., Lin, L., et al.\ 2015, \apj, 804, 125 


\bibitem[Magdis et al.(2016)]{2016MNRAS.456.4533M} Magdis, G.~E., Bureau, M., Stott, J.~P., et al.\ 2016, \mnras, 456, 4533 


\bibitem[Maragkoudakis et al.(2017)]{2017MNRAS.466.1192M} Maragkoudakis, A., Zezas, A., Ashby, M.~L.~N., \& Willner, S.~P.\ 2017, \mnras, 466, 1192 


\bibitem[Mast et al.(2014)]{2014A&A...561A.129M} Mast, D., Rosales-Ortega, F.~F., S{\'a}nchez, S.~F., et al.\ 2014, \aap, 561, A129 


\bibitem[Noeske et al.(2007)]{2007ApJ...660L..43N} Noeske, K.~G., Weiner, B.~J., Faber, S.~M., et al.\ 2007, \apjl, 660, L43 


\bibitem[P{\'e}rez et al.(2013)]{2013ApJ...764L...1P} P{\'e}rez, E., Cid Fernandes, R., Gonz{\'a}lez Delgado, R.~M., et al.\ 2013, \apjl, 764, L1 


\bibitem[Pan et al.(2015)]{2015ApJ...804L..42P} Pan, Z., Li, J., Lin, W., et al.\ 2015, \apjl, 804, L42 


\bibitem[Poggianti \& Barbaro(1997)]{1997A&A...325.1025P} Poggianti, B.~M., \& Barbaro, G.\ 1997, \aap, 325, 1025 
 

\bibitem[S{\'a}nchez et al.(2012)]{2012A&A...538A...8S} S{\'a}nchez, S.~F., Kennicutt, R.~C., Gil de Paz, A., et al.\ 2012, \aap, 538, A8 


\bibitem[S{\'a}nchez et al.(2013)]{2013A&A...554A..58S} S{\'a}nchez, S.~F., Rosales-Ortega, F.~F., Jungwiert, B., et al.\ 2013, \aap, 554, A58


\bibitem[S{\'a}nchez et al.(2016a)]{2016RMxAA..52..171S} S{\'a}nchez, S.~F., P{\'e}rez, E., S{\'a}nchez-Bl{\'a}zquez, P., et al.\ 2016a, \rmxaa, 52, 171 


\bibitem[S{\'a}nchez et al.(2016b)]{2016RMxAA..52...21S} S{\'a}nchez, S.~F., P{\'e}rez, E., S{\'a}nchez-Bl{\'a}zquez, P., et al.\ 2016b, \rmxaa, 52, 21 


\bibitem[S{\'a}nchez et al.(2017)]{2017arXiv170905438S} S{\'a}nchez, S.~F., Avila-Reese, V., Hernandez-Toledo, H., et al.\ 2017, arXiv:1709.05438 


\bibitem[S{\'a}nchez-Menguiano et al.(2016)]{2016A&A...587A..70S} S{\'a}nchez-Menguiano, L., S{\'a}nchez, S.~F., P{\'e}rez, I., et al.\ 2016, \aap, 587, A70 


\bibitem[Salim et al.(2007)]{2007ApJS..173..267S} Salim, S., Rich, R.~M., Charlot, S., et al.\ 2007, \apjs, 173, 267 


\bibitem[Salpeter(1955)]{1955ApJ...121..161S} Salpeter, E.~E.\ 1955, \apj, 121, 161 


\bibitem[Simard et al.(2011)]{2011ApJS..196...11S} Simard, L., Mendel, J.~T., Patton, D.~R., Ellison, S.~L., \& McConnachie, A.~W.\ 2011, \apjs, 196, 11


\bibitem[Speagle et al.(2014)]{2014ApJS..214...15S} Speagle, J.~S., Steinhardt, C.~L., Capak, P.~L., \& Silverman, J.~D.\ 2014, \apjs, 214, 15


\bibitem[Spindler et al.(2018)]{2018MNRAS.476..580S} Spindler, A., Wake, D., Belfiore, F., et al.\ 2018, \mnras, 476, 580


\bibitem[Stoehr et al.(2008)]{2008ASPC..394..505S} Stoehr, F., White, R., Smith, M., et al.\ 2008, Astronomical Data Analysis Software and Systems XVII, 394, 505 


\bibitem[Tacchella et al.(2015)]{2015Sci...348..314T} Tacchella, S., Carollo, C.~M., Renzini, A., et al.\ 2015, Science, 348, 314


\bibitem[Tacchella et al.(2016a)]{2016MNRAS.457.2790T} Tacchella, S., Dekel, A., Carollo, C.~M., et al.\ 2016, \mnras, 457, 2790 


\bibitem[Tacchella et al.(2016b)]{2016MNRAS.458..242T} Tacchella, S., Dekel, A., Carollo, C.~M., et al.\ 2016, \mnras, 458, 242 


\bibitem[Wang et al.(2017a)]{2017ApJ...844..144W} Wang, E., Kong, X., Wang, H., et al.\ 2017a, \apj, 844, 144 


\bibitem[Wang et al.(2017b)]{2017arXiv171007569W} Wang, E., Li, C., Xiao, T., et al.\ 2017b, arXiv:1710.07569 


\bibitem[Wuyts et al.(2013)]{2013ApJ...779..135W} Wuyts, S., F{\"o}rster Schreiber, N.~M., Nelson, E.~J., et al.\ 2013, \apj, 779, 135 


\end{thebibliography}
\end{document}